\documentclass[aps,pra,reprint,10pt,longbibliography]{revtex4-1}

\usepackage{amsmath,amsthm,amsfonts,amssymb,bm}
\usepackage{mdframed,graphicx,array}
\definecolor{OceanBlue}{rgb}{0,0.35,0.7} 
\usepackage{hyperref}
\hypersetup{colorlinks=true,allcolors=OceanBlue}

\newcommand{\pars}[1]{\left(#1\right)} 
\newcommand{\bracs}[1]{\left[#1\right]} 
\newcommand{\set}[1]{\left\{#1\right\}} 

\newcommand{\mr}{\mathrm} 
\newcommand{\mb}{\mathbf} 
\newcommand {\cl}{\mathcal} 
\newcommand {\tsf} [1]{\textsf{#1}} 
\newcommand{\vectsym}{\boldsymbol}  
\DeclareMathOperator{\Tr}{Tr\,}            

\newcommand*\diff{\mathop{}\!\mathrm{d}}

\newcommand{\balpha}{\vectsym{\alpha}}  

\newcommand {\ket}[1] {\left|{#1}\right\rangle}
\newcommand {\kets} [1] {\left|#1\right\rangle_{S}}

\newcommand {\keti} [1] {\left|#1\right\rangle_{I}}

\newcommand {\bra}[1] {\langle{#1}|}

\newcommand{\abs}[1]{\left | #1 \right |}   
\newcommand {\norm} [1] {\left \| #1 \right \|} 

\begin{document}
\title{Fundamental limits of quantum illumination}
\author{Ranjith Nair$^{1,2,}$}
\email{nairanjith@gmail.com}
\author{Mile Gu$^{1,2,3}$}
\affiliation{$^1$ School of Physical and Mathematical Sciences,  \\ Nanyang Technological University, 639673, Singapore}
\affiliation{$^2$ Complexity Institute, Nanyang Technological University, 639673, Singapore}
\affiliation{$^3$ Centre for Quantum Technologies, 117543, Singapore}

\date{\today}




\begin{abstract}
In Quantum Illumination (QI), a signal beam initially entangled with an idler beam held at the receiver interrogates a target region bathed in  thermal background  light. The returned beam is  measured jointly with the idler in order to determine whether a weakly reflecting target is present.  Using tools from quantum information theory, we derive lower bounds on the average error probability of detecting both specular and fading targets and on the mean squared error of estimating the reflectance of a detected target, which are obeyed by any QI transmitter satisfying a signal energy constraint.  For bright thermal backgrounds, we  show that the QI system using multiple copies of low-brightness two-mode squeezed vacuum states is nearly optimal.  More generally, our results place limits on the best possible performance achievable using QI systems at all wavelengths, and at all signal and background noise levels.
\end{abstract}

\maketitle
\noindent
Quantum illumination (QI) is a photonic quantum sensing  protocol (see \cite{PBG+18} for an overview of the field) introduced by Lloyd \cite{Llo08} in which entanglement shared between a signal beam interrogating a target region and a locally held idler beam is used to detect the presence of a weakly reflecting target better than a strategy using only a signal beam of the same energy, i.e., average photon number. Soon after \cite{Llo08} appeared, Tan et al. found  that using multiple signal-idler modes prepared in the two-mode squeezed vacuum (TMSV) state allows a 6 dB improvement of the error probability exponent relative to a classical ladar (laser detection and ranging) system of the same energy \cite{TEG+08}. Surprisingly -- and unlike other sensing applications for which quantum advantage quickly disappears in the presence of decoherence \cite{EdMFD11,DKG12} -- this 6 dB advantage occurs when the target is bathed in  thermal background radiation of brightness (i.e.,  per-mode energy) $N_B \gg 1$, due to which the initial signal-idler entanglement is lost. 
These results have inspired much theoretical \cite{SL09,GE09,BGW+15,SLHG-R+17,ZZS17SFG,ZZS17Rayleighfading,ZZS17ROC,WTL+17,DPB18,WPT+16,YMZ18arxiv,PLL+19} and experimental \cite{LR-BD13,ZTZ+13,ZMW+15,CVB+19,LGH+19,BPV+20} work in QI, much of which is reviewed in \cite{PBG+18,Sha20}. 

In this paper, we use tools from quantum information \cite{NC00}  to investigate the fundamental limits of bosonic QI systems in all regimes of signal and background noise strength and allowing for all possible choices of quantum states at the transmitter and quantum measurements at the receiver. We first derive a lower bound on the average error probability of any QI system that transmits an $M$-mode signal beam entangled with locally held idler modes under a total signal energy constraint $\mathcal{N}_S$. For  $N_B \gg 1$, we show that the  scheme of \cite{TEG+08} using independently and identically distributed (iid) copies of  TMSV states with signal brightness  $N_S \equiv \mathcal{N}_S/M \ll 1$ achieves the greatest error probability exponent allowed by quantum mechanics, and that its near optimality persists for the detection of targets exhibiting flat Rayleigh fading. We also show that any QI system for estimating the reflectance of a detected target has a mean squared error that is at least half that suffered by the best classical ladar.

\section{QI Setup and Background} \label{sec:setup}

\begin{figure}[tbp]
\centering
{\includegraphics[trim=22mm 70mm 22mm 42mm, clip=true,width=\columnwidth]{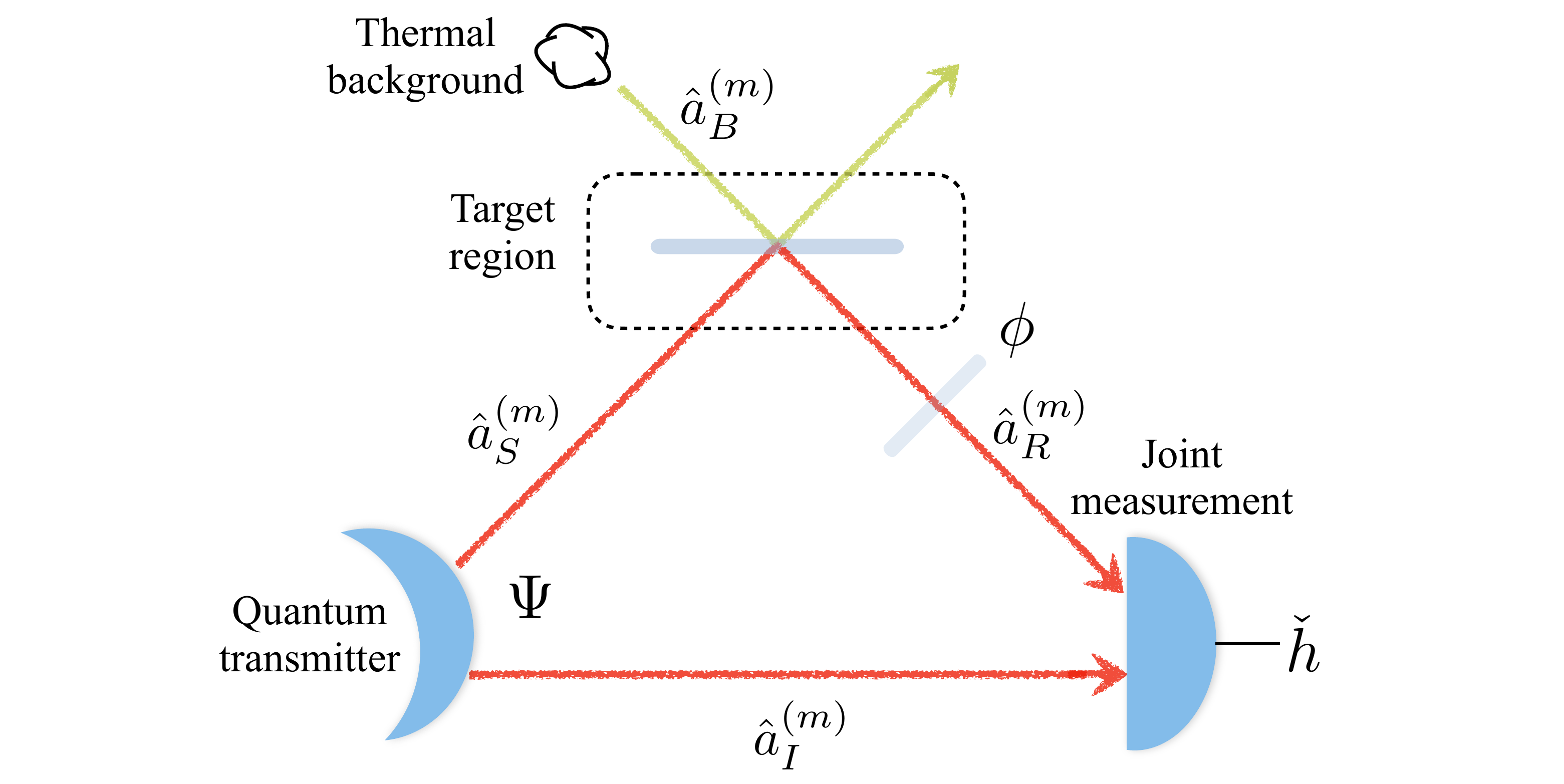}}
\caption{General setup for QI: An entangled state $\Psi$ of $M$ signal and idler modes is prepared. Each signal mode $\hat{a}_S^{(m)}$  interrogates a distant target region  bathed in  thermal radiation $\hat{a}_B^{(m)}$ which may contain a target of effective reflectance $\eta \ll 1$. The modes $\{\hat{a}_R^{(m)}\}_{m=1}^M$ returned from the target region are measured jointly with the unperturbed idler modes $\{\hat{a}_I^{(m)}\}_{m=1}^M$ in order to decide whether or not the target was present. }
\label{fig:QIsetup}
\end{figure}

A general QI setup is depicted schematically in Fig.~\ref{fig:QIsetup}. A transmitter prepares an entangled quantum state of $M$ signal (denoted $S$) modes (annihilation operators $\{\hat{a}_S^{(m)}\}_{m=1}^M$) and idler (denoted $I$) modes (annihilation operators $\{\hat{a}_I^{(m)}\}_{m=1}^{M}$). The signal modes interrogate a target region which may contain a weakly reflecting target  bathed in background (denoted $B$) light (annihilation operators $\{\hat{a}_B^{(m)}\}_{m=1}^M$). The annihilation operators $\{\hat{a}_R^{(m)}\}_{m=1}^M$ of the modes returned from the target region, which may acquire an additional phase shift $\phi$, obey the relation
\begin{align} \label{HP}
\hat{a}_R^{(m)} &=  \sqrt{\eta_h}\,e^{i\phi}\, \hat{a}_S^{(m)} + \sqrt{1-\eta_h} \,\hat{a}_B^{(m)}.
\end{align} 
Here $h=0 \,(1)$ indicates the absence (presence) of a target, $\eta_0 = 0$, and $\eta_1 = \eta \ll 1$ is the effective reflectance of the  signal to return beam path when the target is present. The two hypotheses are assigned prior probabilities $\{\pi_h\}_{h=0}^1$. In order to derive fundamental physical limits, we assume initially that the target is specular and that the values of $\eta$ and $\phi$ are known to the receiver. As in previous works \cite{TEG+08,SL09,GE09,BGW+15,SLHG-R+17,ZZS17SFG,ZZS17Rayleighfading,ZZS17ROC,WTL+17,DPB18,Sha20}, each  background mode is assumed to be in a thermal state $\rho_{\tsf{th}}(N_B^{(h)}) = \sum_{n=0}^\infty {N_B^{(h)}}^n/ {\pars{N_B^{(h)} +1 }}^{n+1} \ket{n}\bra{n}_B$ ($\{\ket{n}_B\}$ are number states of $B$) of brightness 
\begin{align} \label{NBh}
\big\langle{\hat{a}_B^{(m)^\dag} \hat{a}_B^{(m)}}\big\rangle_h \equiv N_B^{(h)} = N_B/(1-\eta_h)
\end{align}
under  hypothesis $h$, where $N_B$ is a given nominal value of the background brightness. ~Eq.~\eqref{NBh} ensures that the target leaves no passive signature, and cannot be detected if a vacuum state is transmitted. Since $\eta \ll 1$ for standoff sensing, we have $N_B^{(1)} \simeq N_B = N^{(0)}_B$.

Let $\ket{\mb{n}}_S$ denote a product number state of the $M$ signal modes with $\mb{n} = (n_1,\ldots, n_M)$. The most general QI strategy  consists of preparing a pure quantum state 
\begin{align} \label{transmitter}
\ket{\psi}_{IS} &= \sum_{\mb{n}} \sqrt{p_\mb{n}} \keti{\chi_\mb{n}}\kets{\mb{n}},
\end{align} 
of the $IS$ system (called  \emph{transmitter} hereafter) subject only to the total signal energy constraint
$\sum_{\mb{n}} p_\mb{n} (n_1 + \cdots + n_M) \equiv \sum_{n=0}^\infty n\, p_n = \cl{N}_S.$
 Here,  $\set{\keti{\chi_\mb{n}}}$ is any normalized (not necessarily orthogonal) set of idler states and $\set{p_n}$ is the probability mass function of the \emph{total} photon number $n= \sum_{m=1}^M n_m$ in the $S$ modes. 
In the Schr\"odinger picture, the evolution (\ref{HP}) corresponds to the output density operators 
\begin{align} \label{dos}
\rho_h &= \Big[{\mr{id}_I \otimes {\big( \cl{U}_{\phi} \circ \cl{L}_{\eta_h,N_B^{(h)}}\big)}^{\otimes M} }\Big]\pars{\Psi},
\end{align}
where $\Psi = \ket{\psi}\bra{\psi}_{IS}$, $\mr{id}_I$ denotes the identity channel on $I$,  and unitary phase-shift channels $\cl{U}_\phi$ and  \emph{noisy attenuator} channels $\cl{L}_{\eta_h, N_B^{(h)}}$ of transmittance $\eta_h$ and added noise $N_B^{(h)}$ (see Fig.~\ref{fig:channeldecomp}, \cite{CGH06,G-PN-BL+12})  act on each mode of $S$. Note that $\rho_0 = \pars{\Tr_S \Psi} \otimes \rho_{\tsf{th}}(N_B)$ for any value of $\phi$, where $\Tr_S$ indicates  partial trace over $S$.

The receiver makes a joint measurement with outcome $\check{h} \in \{0,1\}$ on the returned and idler modes  that minimizes the average error probability
$P_e\bracs{\rho_0,\rho_1} = \pi_0 P_F + \pi_1 P_M,$
where $P_F = \mr{Pr}[\check{h}=1 | h=0]$ is the false-alarm probability and $P_M = \mr{Pr}[\check{h}=0 | h=1]$ is the miss probability. The lowest achievable $P_e$ is given by the \emph{Helstrom limit} \cite{Hel76}:
\begin{align} \label{Helstrom}
P_e\bracs{\rho_0,\rho_1}  = 1/2 - \norm{\pi_0 \rho_o - \pi_1 \rho_1}_1/2,
\end{align}
where $\norm{X}_1 = \Tr \sqrt{X^\dag X}$ is the operator trace norm. Since the trace norm is hard to calculate, we often resort to bounds on it. 

To be useful, a  QI system must have a lower $P_e$ than the best classical ladar, i.e., a transmitter that is in a coherent state or a mixture of product signal-idler coherent states. In \cite{TEG+08}, it was shown that no classical ladar with signal energy $\cl{N}_S$ can have an error probability lower than
 \begin{align} \label{CI}
 P_e^{\tsf{cl}} \simeq \sqrt{\pi_0 \pi_1}\, \exp\Big[{-\eta\cl{N}_S \Big({\sqrt{N_B+1}- \sqrt{N_B}}}\Big)^2\Big].
 \end{align}
 In particular, the exponent after the minus sign in the above expression is the best possible for a classical ladar and  $\simeq \eta \cl{N}_S/ 4N_B$ in the regime $N_B \gg 1$. It was also shown using the quantum Chernoff bound \cite{ACM+07,PL08} that  if  $M$ copies of the TMSV state $\ket{\psi}_{I_m S_m} = \sum_{n=0}^{\infty} \sqrt{N_S^n/\pars{N_S+1}^{n+1}} \ket{n}_{I_m}\ket{n}_{S_m}$ are transmitted, there exists a measurement that achieves
 \begin{align} \label{TMSVQI}
 P_e^{\tsf{TMSV}} \simeq \sqrt{\pi_0 \pi_1}\, \exp{(-\eta\cl{N}_S/N_B)}
 \end{align}
 in the regime of signal brightness $N_S = \cl{N}_S/M \ll 1$ and noise brightness $N_B \gg 1$. Recently, a concrete receiver using sum frequency generation (SFG) was proposed \cite{ZZS17SFG} that can in principle realize this 6 dB error exponent advantage over the classical performance (\ref{CI}).
  
\section{Methods and Results} \label{sec:methods}

\begin{figure}[tbp]
\centering
{\includegraphics[trim=14mm 80mm 10mm 75mm, clip=true,width=\columnwidth]{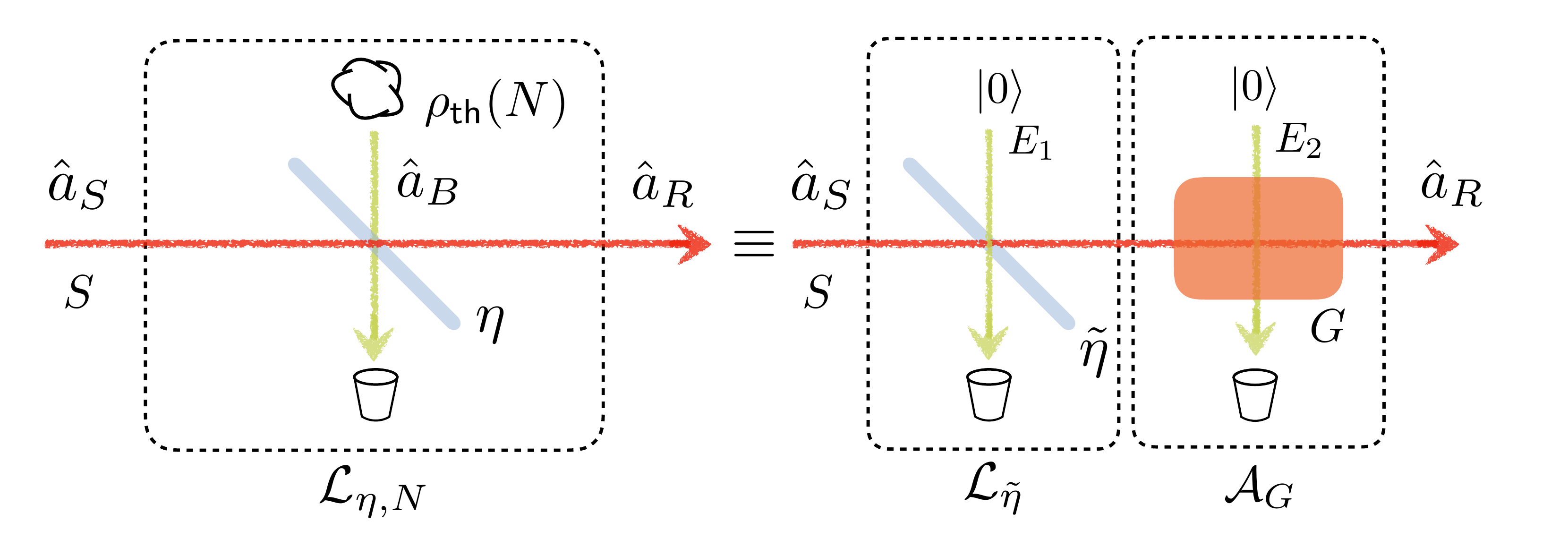}}
\caption{Left: A noisy attenuator channel $\cl{L}_{\eta, N}$  mixes the signal mode $S$ with a background mode $B$ in a thermal state of energy $N$ at a beam splitter of transmittance $\eta$. Right:  $\cl{L}_{\eta, N}$ can be realized as a cascade of a quantum-limited (mode $E_1$  in vacuum) loss channel $\cl{L}_{\widetilde{\eta}}$ with $\widetilde{\eta} = \eta/G$ and a quantum-limited (mode $E_2$ in vacuum) amplifier (two-mode squeezer) $\cl{A}_G$ of gain $G= (1-\eta)N +1$. }
\label{fig:channeldecomp}
\end{figure}

In this paper, we are interested in performance bounds valid for arbitrary transmitters $\Psi$. A key tool that we use is the decomposition of the noisy attenuator channels appearing in Eq.~\eqref{dos} in terms of quantum-limited attenuators and amplifiers. Specifically, we can write $\cl{L}_{\eta, N}$ as the concatenation (see Fig.~\ref{fig:channeldecomp}) 
\begin{align} \label{decompositionparameters}
\cl{L}_{\eta,N} &= \cl{A}_{{G}} \circ \cl{L}_{\widetilde{\eta}},
\end{align}
 where $\cl{L}_{\widetilde{\eta}} := \cl{L}_{\widetilde{\eta},0}$, $\cl{A}_G$ is a quantum-limited amplifier of gain $G= (1-\eta)N + 1$, and $\widetilde{\eta} = {\eta}/G$ \cite{CGH06,G-PN-BL+12}.

We also make much use of the \emph{fidelity} between two  states $\rho$ and $\sigma$ of a quantum system  defined as $F(\rho,\sigma) = \Tr \sqrt{\sqrt{\rho} \,\sigma \sqrt{\rho}}$, which is an important measure of closeness between quantum states \cite{NC00}. The fidelity satisfies the \emph{data processing inequality}  $F(\cl{C}(\rho),\cl{C}(\sigma))  \geqslant F(\rho,\sigma)$, where $\cl{C}$ is any quantum channel \cite{NC00}.

\subsection{Lower bound on QI error probability} \label{sec:Pelb}

We begin by using  Eqs.~(\ref{NBh}) and (\ref{decompositionparameters}) to write
\begin{align} \label{QIparams}
\cl{L}_{\eta_h,N_B^{(h)}} &= \cl{A}_{N_B + 1} \circ \cl{L}_{\eta_h/(N_B + 1)} \equiv \cl{A}_{N_B + 1} \circ \cl{L}_{\widetilde{\eta}_h}.
\end{align}
Then, using the data processing inequality on the states in Eq.~\eqref{dos} gives
\begin{align}
 F\pars{\rho_0, \rho_1} &\geqslant
 F\pars{\Big[{\mr{id}_I \otimes \cl{L}_{\widetilde{\eta}_0}^{\otimes M} }\Big] ({\Psi}), \bracs{\mr{id}_I \otimes \cl{L}_{\widetilde{\eta}_1}^{\otimes M} } (\Psi) } \\
 &\geqslant \sum_n p_n \mu^n,
\end{align}
where  the last inequality follows from the result of \cite{Nai11} (Sec.~II) with $\mu = \sqrt{1-\eta/(N_B+1)}$. Using the inequalities $P_e[\sigma_0,\sigma_1] \geqslant \pars{1 - \sqrt{1-4\pi_0\pi_1F^2\pars{\sigma_0, \sigma_1}}}/2 \geqslant \pi_0\pi_1F^2\pars{\sigma_0, \sigma_1} $ relating the Helstrom limit Eq.~\eqref{Helstrom} and the fidelity  for any two states $\sigma_0$ and $\sigma_1$ \cite{FvdG99,Aud14}, we get the bound
\begin{align} \label{PelbPsi}
P_e^{\Psi} \geqslant \pi_0\pi_1\bigg\{\sum_{n=0}^\infty p_n \big[1-{\eta}/({N_B+1})\big]^{n/2}\bigg\}^2 
\end{align}
on the $P_e$ of any  $\Psi$. Further, using Jensen's inequality on the convex function $x \mapsto \mu^x$  gives the $\Psi$-independent bound
\begin{align} \label{Pelb}
P_e^{\tsf{QI}} \geqslant \pi_0 \pi_1  \exp\pars{-\beta \cl{N}_S},
\end{align}
where we have defined the exponent $\beta:= -\ln[1-\eta/(N_B+1)]$. 

Eqs.~(\ref{PelbPsi})-(\ref{Pelb}) make up our first result. Eq.~\eqref{Pelb} shows that no QI system with signal energy $\cl{N}_S$ can have an error probability exponent greater than $\beta \cl{N}_S$. In the low-noise limit $N_B \simeq 0$, we have  $\beta \simeq \eta$, and the best possible exponent $ \simeq \eta\cl{N}_S$. This matches that achieved by a classical ladar (Eq.~\eqref{CI}) and is consistent with earlier no-go results for QI advantage in this regime  \cite{Nai11,SL09}. In the high-noise regime $N_B \gg 1$, the optimum exponent  $ \simeq \eta \cl{N}_S/N_B$ is attained by the TMSV QI system \cite{TEG+08} (cf. Eq.~\eqref{TMSVQI}). 

Eq.~\eqref{PelbPsi} explains why a TMSV QI system  must use a large $M$ in order to beat the classical performance  given by Eq.~\eqref{CI} for any $N_B$. Intuitively, the right-hand side of Eq.~\eqref{PelbPsi} decreases the more the  distribution $\left\{p_n \right\}$ is concentrated around its mean $\cl{N}_S$. For TMSV QI, the variance of the total signal photon number is $\cl{N_S}\pars{N_S +1}$, which is minimized  as the brightness $N_S \rightarrow 0$ for fixed $\cl{N}_S$. More precisely, evaluating Eq.~\eqref{PelbPsi} gives:
\begin{align} \label{Pelbtmsv} 
P_e^{\tsf{TMSV}} \geqslant \pi_0 \pi_1 \bigg[1 + \cl{N}_S \Big(1 - \sqrt{1 - \eta/(N_B+1)}\Big)/M\bigg]^{-2M}. 
\end{align}
Thus, for  fixed $M$, the error probability of TMSV QI cannot decay exponentially  with $\cl{N}_S$. However, if $M \rightarrow \infty$ for fixed $\cl{N}_S$, Eq.~\eqref{Pelbtmsv} allows the exponential scaling of Eq.~\eqref{Pelb}.

\subsection{Targets exhibiting flat Rayleigh fading} \label{sec:fading}
 
 At optical wavelengths, most target surfaces are rough and our model of Eq.~\eqref{HP} involving  deterministic and known values for $\eta$ and $\phi$ needs modification. Such targets often obey a \emph{Rayleigh fading} model \cite{VanTreesI,Sha82} where  $\eta$ and $\phi$  are independent random variables with $\eta$ distributed according to an exponential probability density $P(\eta)$ of mean $\overline{\eta}$ (the average reflectance of the target) and  $\phi$ uniformly distributed over $[0,2\pi)$. In addition, we  consider the \emph{flat fading} limit in which the values of $\eta$ and $\phi$ do not vary over the $M$ signal modes. Such a model  was used in the TMSV QI study \cite{ZZS17Rayleighfading}.  With these assumptions, $\rho_1$ in Eq.~\eqref{dos} is replaced by
 \begin{align} \label{fadingrho1}
 \rho_1 = \frac{1}{2\pi}\int_0^1 \diff \eta \,P(\eta) \int_{0}^{2\pi} \diff \phi  \Big[{\mr{id}_I \otimes {\big(\cl{U}_{\phi} \circ \cl{L}_{\eta,N_B^{(\eta)}} \big)}^{\otimes M} }\Big] \pars{\Psi}
 \end{align}
 while $\rho_0$ remains the same as before. In Eq.~\eqref{fadingrho1}, $N_B^{(\eta)} = N_B/(1-\eta)$ can vary greatly in the region $\eta \approx 1$, but the model is an excellent  one if $\overline{\eta} \ll 1$, as is the case in practice.

 With some more work (see Appendix \ref{sec:appA} for details), the  approach of Sec.~\ref{sec:Pelb} can be extended to give the  lower bound
 \begin{align} \label{Rayleighfadingbound}
 P_e^{\tsf{QI}\mr{; fading}} \geqslant \pi_0\pi_1/\bracs{1 + \overline{\eta}\cl{N}_S \ln \pars{1 +1/N_B}}
\end{align}
for {any} transmitter with signal energy $\cl{N}_S$. We see that for fading targets the error probability no longer decays exponentially with $\cl{N}_S$, generalizing the result of \cite{ZZS17Rayleighfading} beyond just TMSV QI in the  $N_S \ll 1, N_B \gg 1$ regime. In the same regime, it was shown that the SFG  receiver of \cite{ZZS17SFG} can in principle achieve $ P_e^{\tsf{TMSV}\mr{; fading}} \simeq \pi_1/\pars{1 + \overline{\eta}\cl{N}_S/N_B}$ for $\overline{\eta}\cl{N}_S/N_B \gg 1$, while  the best classical transmitter has a $P_e$ that is greater by a factor  $\simeq \ln \pars{\overline{\eta}\cl{N}_S/N_B}$ \cite{ZZS17Rayleighfading}. Comparing with Eq.~\eqref{Rayleighfadingbound}, we see that TMSV transmitters with SFG reception perform essentially optimally among all transmitters with the same $\cl{N}_S$.

\subsection{Estimation of target reflectance} \label{sec:estimation}

Finally, we consider the problem of \emph{estimating} the effective reflectance $\eta \ll 1$ of a detected  specular target. The setup of Fig.~\ref{fig:QIsetup} applies  except that a target is known to be present and the joint  measurement of the returned signal and idler modes generates an estimate $\check{\eta}$ of  $\eta$. Instead of Eq.~\eqref{dos}, we  have a family of density operators $\left\{\rho_\eta\right\}$ given by
\begin{align} \label{estimationdos}
\rho_{\eta} &= \Big[\mr{id}_I \otimes \big(\cl{U}_{\phi} \circ \cl{L}_{\eta,N_B^{(\eta)}} \big)^{\otimes M} \Big] \pars{\Psi},
\end{align}
where, as before, $\Psi$ is a general energy-constrained $M$-mode transmitter, $N_B^{(\eta)} = N_B/(1-\eta)$, and $\phi$ is assumed to be known.
The mean squared error $\mr{MSE}_{\eta} = \mathbb{E} \bracs{\pars{\check{\eta} - \eta}^2}$ of any unbiased estimator $\check{\eta}$ of $\eta$ obeys the quantum Cram\'er-Rao bound \cite{Hel76,PBG+18}:
\begin{align} \label{QCRB}
\mr{MSE}_{\eta} \geqslant \cl{K}_{\eta}^{-1} = \Big[ - 4 \,{\partial^2 F \big(\rho_{\eta}, \rho_{\eta'}\big)}/{\partial \eta'^2} \big\vert_{\eta' = \eta}\Big]^{-1},
\end{align}
where $\cl{K}_\eta$ is the \emph{quantum Fisher information} (QFI) on $\eta$, and the equality relating QFI to fidelity is due to  \cite{BC94}. 

Using  data processing arguments similar to those used for our target detection results, we can derive the upper bound
\begin{align} \label{QFIub}
\cl{K}_\eta^{\tsf{QI}} \leqslant \cl{N}_S/\bracs{\eta (N_B +1 -\eta)}
\end{align}
on the QFI valid for any $\Psi$, and the upper bound
\begin{align} \label{QFIcl}
\cl{K}_\eta^{\tsf{cl}} \leqslant \cl{N}_S/\bracs{\eta (2N_B +1)}
\end{align}
on the QFI of any classical transmitter  which is achieved using coherent states (see Appendix \ref{sec:appB} for details).   For $N_B=0$, Eq.~\eqref{QFIub} recovers the result of \cite{Nai18loss}  that holds for all $\eta$ and is achieved by a large class of transmitters. For $N_B >0$ and $\eta \ll 1$, it is  close to an upper bound on the QFI  derived in \cite{SLHG-R+17}, though only for iid transmitters. In the presence of excess noise, Eq.~\eqref{QFIub} need not be saturable, but it was shown ( Eq.~(6) of \cite{SLHG-R+17}\footnote{Ref.~\cite{SLHG-R+17} reports the per-mode QFI $\cl{K}_{\sqrt{\eta}}/M$ on the {reflectivity} $\sqrt{\eta}$, which is related to  $\cl{K}_{\eta}$ via  $\cl{K}_{\eta} = \pars{\partial \sqrt{\eta}/\partial \eta}^2 \cl{K}_{\sqrt{\eta}}$.}) that the QFI of an $M$-mode TMSV transmitter is 
\begin{align} \label{QFItmsv}
\cl{K}_{\eta}^{\tsf{TMSV}} = \bracs{\cl{N}_S (1+N_S)}/\bracs{\eta\pars{2N_SN_B + N_S + N_B +1}}.
\end{align}
In the low-brightness limit $N_S \rightarrow 0$, Eq.~\eqref{QFItmsv} approaches the bound Eq.~\eqref{QFIub} for all values of $N_B$ when $\eta \ll 1$. Moreover, as argued in \cite{SLHG-R+17}, this bound is achieved by the optical parametric amplifier receiver of \cite{GE09} (see also \cite{SZZ20}).  From Eqs.~(\ref{QCRB})-(\ref{QFItmsv}), we see  that the advantage of using quantum transmitters in the $\eta \ll 1$ regime is limited to at best a factor of 2  in the MSE.

\section{Discussion and Outlook} \label{sec:discussion}
Our study of QI target detection took place in the Bayesian paradigm in which prior probabilities are assigned to the two hypotheses. Target detection is often treated in the alternative Neyman-Pearson setting \cite{Hel76} where one asks for the best achievable detection probability $P_D = 1 - P_M$ for given $P_F$ (which determines the so-called Receiver Operating Characteristic (ROC)). TMSV QI has been extensively studied  in this paradigm \cite{WTL+17,ZZS17ROC,Sha20}, and shown to be asymptotically optimal in a specific sense in \cite{DPB18}. We note that the fidelity lower bounds developed here can also be used to place upper bounds on the ROC  curve  of arbitrary QI transmitters (see, e.g., \cite{TN12,Tsa13}). We can also ask if adaptively chosen transmitter states help in QI target detection -- some suggestive results are known for the  Bayesian \cite{PLL+19} and Neyman-Pearson \cite{CMW16,BHK+18arxiv} paradigms.

Our reflectance estimation results show that the gain in using quantum transmitters is modest when $\eta \ll 1$. It is therefore interesting to explore quantum limits in detection and estimation in the region of moderate $\eta$ and $N_B$, which is relevant to detection applications like quantum reading \cite{Pir11} and various estimation applications (see, e.g., \cite{Nai18loss} and references therein). Our techniques may also be extended to protocols of phase-randomized quantum illumination without amplitude fading,  e.g., the implementations \cite{LR-BD13,LGH+19}.

While further study of operational limitations and nonidealities in QI systems is necessary to assess their performance in real-world situations \cite{Sha20}, our work  delineates their ultimate capabilities for all ranges of system parameters. We also believe that the  tools developed here may help to study quantum limits of other optical sensing and communication protocols operating in noisy high-loss conditions. 

%
%

\section*{Funding Information}
National Research Foundation Singapore (NRF-NRFF2016-02); NRF Singapore and L'Agence Nationale de la Recherche Joint Project (NRF2017-NRFANR004 VanQuTe); Singapore Ministry of Education (MOE2019-T1-002-015); FQXi (FQXi-RFP-IPW-1903).

\section*{Acknowledgments}

We thank Nicolas Menicucci, Jeffrey Shapiro, and Quntao Zhuang for helpful comments.

\onecolumngrid
\appendix
\section{ Targets exhibiting flat Rayleigh fading: Error probability lower bound for QI} \label{sec:appA}

We asserted in the main text that, for targets exhibiting flat Rayleigh fading, the density operators of the joint return-idler system when the target is absent and present are given by
\begin{align} \label{fadingdos}
\rho_0 &= \bracs{\mr{id}_I \otimes \pars{ \cl{L}_{0,N_B}^{\otimes M}} }\pars{\Psi},  \\
 \rho_1 &= (1/2\pi)\int_0^1 \diff \eta \,P(\eta) \int_{0}^{2\pi} \diff \phi  \bracs{\mr{id}_I \otimes \pars{\cl{U}_{\phi} \circ \cl{L}_{\eta,N_B^{(\eta)}}}^{\otimes M} }\pars{\Psi} \label{fadingrho1}
 \end{align}
 respectively.  It is usual in the classical radar literature to assume that $\sqrt{\eta}$ has a Rayleigh distribution -- see, e.g., Sec.~4.4.2 of \cite{VanTreesI}.  Then $\eta$ itself has the exponential probability density $\widetilde{P}(\eta) = \pars{1/\overline{\eta}} \exp\pars{-\eta/\overline{\eta}}$ supported on $ \eta \geqslant 0$. Strictly speaking, the probability that $\eta > 1$ should be zero since the target is a passive reflector. However, the above model is an excellent approximation for a diffuse reflector as long as $\overline{\eta} \ll 1$, which is usually the case in practice.
 
 Quantum mechanically, however, Eq.~(1) of the main text does not represent a physically possible transformation if $\eta >1$. To deal with this issue, we replace $\widetilde{P}(\eta)$ with the truncated exponential density
 \begin{align} \label{P}
 P(\eta) =\left\{
	\begin{array}{ll}
		\exp\pars{-\eta/\overline{\eta}} /\bracs{\overline{\eta}\pars{1 - e^{-1/\overline{\eta}}}}  & \mbox{if } \eta \in[0,1]  \\
		0 & \mbox{if } \eta \geqslant 1.
	\end{array}
\right.
\end{align}
Again, if $\overline{\eta} \ll 1$, the discrepancy between Eq.~\eqref{P} and  $\widetilde{P}(\eta)$  is negligible. It is the probability density of Eq.~\eqref{P} that appears in Eq.~\eqref{fadingrho1} and Eq.~(15) of the main text. Finally, note that setting $N_B^{(\eta)} = N_B/\pars{1-\eta}$ in Eq.~\eqref{fadingrho1} enforces the no-passive-signature assumption in this fading scenario. While this implies that $N_B^{(\eta)}$ can vary greatly in the vicinity of $\eta \approx 1$, such large deviations of the background noise in the model have very low probability  if $\overline{\eta} \ll 1$.

We can now proceed to develop our error probability lower bound. First, we observe that the squared fidelity $F^2(\rho,\sigma)$, like $F(\rho,\sigma)$ itself \cite{NC00}, is concave in each of its arguments \cite{Joz94}, so that we can write
\begin{equation}
\begin{aligned}
F^2(\rho_0,\rho_1) &\geqslant  \pars{1/2\pi} \int_0^1 \diff \eta \,P(\eta) \int_{0}^{2\pi} \diff \phi  \\
& \times F^2 \left\{\rho_0, \bracs{\mr{id}_I \otimes \pars{\cl{U}_{\phi} \circ \cl{L}_{\eta,N_B^{(\eta)}}}^{\otimes M} }\pars{\Psi}\right\}.
\end{aligned}
\end{equation}
Noting that the fidelity appearing in the integrand is $\phi$-independent, we can apply the inequalities of Eqs.~(10)-(11) of the main text to it and use the bound $P_e\bracs{\sigma_0,\sigma_1} \geqslant \pi_0\pi_1\,F^2\pars{\sigma_0,\sigma_1}$  to get the lower bound
\begin{align} \label{PsiPebound}
P_e^{\Psi\mr{;fading}} \geqslant \pi_0 \pi_1 \int_0^1 \diff \eta \,P(\eta) \bracs{\sum_{n=0}^\infty p_n \pars{1 - \frac{\eta}{N_B +1}}^{n/2} }^2
\end{align}
on the average error probability of detecting a fading target.
For any given transmitter $\Psi$ with corresponding $\left\{p_n \right\}$, the right-hand side can be evaluated analytically in some cases, and numerically otherwise.

We can further derive an analytical transmitter-independent bound as follows. Applying Jensen's inequality to the quantity in brackets in Eq.~\eqref{PsiPebound} gives
\begin{align} \label{fidbound1}
P_e^{\Psi\mr{;fading}}   &\geqslant \pi_0 \pi_1\int_0^1 \diff \eta \,P(\eta) \pars{1 - \frac{\eta}{N_B +1}}^{\cl{N}_S}.
 \end{align}
 For $N_B > 0$ and $0 \leqslant \eta \leqslant 1$,  we have $1 - \eta/(N_B +1) \geqslant \exp(- \gamma \eta)$, where $\gamma = \ln(1 +1/N_B)$  is chosen such that the graph of $\exp(-\gamma \eta)$ intersects that of $1 - \eta/(N_B +1)$ at $\eta=0$ and $\eta=1$. Substituting this lower bound into Eq.~\eqref{fidbound1} and evaluating the integral  gives
 \begin{align} \label{fidbound}
 P_e^{\tsf{QI}\mr{;fading}} &\geqslant {\pi_0\pi_1}\frac{1 - \exp\pars{- \gamma \cl{N}_S - 1/\overline{\eta}}} { \bracs{1 - \exp\pars{-1/\overline{\eta}}} \pars{1 + \overline{\eta}\gamma\cl{N}_S}},\\
&\geqslant \frac{ \pi_0\pi_1}{ 1 + \overline{\eta}\gamma\cl{N}_S},
 \end{align}
which is Eq.~(16) of the main text.

\section{Estimation of target reflectance} \label{sec:appB}

In this section, we provide derivations of the results pertaining to estimating the reflectance $\eta \ll 1$ of a weakly reflecting specular target.  As described in the main text, for any transmitter $\Psi$, the density operator $\rho_\eta$ of the returned signal and idler modes conditioned on the target reflectance having the value $\eta$ is given by
\begin{align} \label{estimationdos}
\rho_{\eta} &= \bracs{\mr{id}_I \otimes \pars{\cl{U}_{\phi}^{\otimes M} \circ \cl{L}_{\eta,N_B^{(\eta)}}^{\otimes M}} }\pars{\Psi}, \\
&= \bracs{\mr{id}_I \otimes \pars{\cl{U}_{\phi}^{\otimes M} \circ  \cl{A}_{N_B+1}^{\otimes M} \circ \cl{L}_{\eta/(N_B + 1)}^{\otimes M}} }\pars{\Psi}, \label{dp2}
\end{align}
where we have used the decomposition of Eq.~(8) of the main text. Now note that the quantum channel $\cl{U}_{\phi}^{\otimes M} \circ  \cl{A}_{N_B+1}^{\otimes M}$ that is applied `downstream' to the $S$ system is $\eta$-independent, and can be realized by coupling an  ancilla mode $A$ in a fixed state to the $S$ system and evolving the joint system under a fixed unitary (this is the so-called Stinespring dilation of a quantum channel \cite{NC00}). The monotonicity property of the QFI under partial trace \cite{Pet08qits} then implies that the QFI on $\eta$ achieved by making a measurement on the joint $ISA$ system is at least as much as that on the $IS$ system alone. On the other hand, the invariance of QFI under a known $\eta$-independent unitary transformation implies that the former value equals the QFI on $\eta$ of the state family
\begin{align} \label{purifiedestimationdos}
\sigma_{\eta} &= \bracs{\mr{id}_I \otimes  \cl{L}_{\eta/(N_B + 1)}^{\otimes M} }\pars{\Psi}. 
\end{align}

We have thus reduced the problem to maximizing the QFI on $\eta$ for the outputs $\left\{\sigma_{\eta}\right\}$ of \emph{pure-loss} channels under an energy constraint on the $S$ modes. This problem was solved in \cite{Nai18loss} (cf. Eq.~(14) therein), and transforming variables in that result gives the upper bound
\begin{align}
\cl{K}_{\eta}^{\tsf{QI}} \leqslant \frac{\cl{N}_S}{\eta\pars{N_B +1 -\eta}}
\end{align}
for the QFI of any transmitter $\Psi$ for any value of the excess noise $N_B$, which reproduces Eq.~(19) of the main text.

Consider a single-mode coherent-state transmitter $\kets{\psi} = \kets{\sqrt{\cl{N}_S}}$ of energy $\cl{N}_S$. In order to evaluate the QFI on $\eta$, we first calculate the fidelity between the states $\rho_\eta^{\tsf{CS}}$ and $\rho_{\eta'}^{\tsf{CS}}$ of Eq.~\eqref{estimationdos} for any two values  $\eta$ and $\eta'$. Using known results on the fidelity between Gaussian states (see e.g., Eq.~(3.7) of \cite{MM07}), we have
\begin{align} \label{CSfid}
F\pars{\rho_\eta^{\tsf{CS}},\rho_{\eta'}^{\tsf{CS}}} = \exp\bracs{-\frac{\pars{\sqrt{\eta'} - \sqrt{\eta}}^2\cl{N}_S} {4N_B +2} }
\end{align}
The QFI then follows as
\begin{align} \label{CSQFI}
\cl{K}_{\eta}^{\tsf{CS}} &= - 4 \frac{\partial^2 F\pars{\rho_\eta^{\tsf{CS}},\rho_{\eta'}^{\tsf{CS}}}} {\partial \eta'^2} \Bigg\vert_{\eta' = \eta} \nonumber \\
&= \frac{\cl{N}_S}{\eta(2 N_B +1)}.
\end{align}
The additivity of the QFI for product states \cite{Pet08qits} and the linearity of the coherent-state QFI (\ref{CSQFI}) in the energy imply that (\ref{CSQFI}) is also the QFI of a multimode coherent state of total energy $\cl{N}_S$. Finally, any classical-state transmitter  can be written as a proper $P$-representation \cite{MW95}, i.e., in the form 
\begin{align} \label{classical probe}
\rho = \int_{\mathbb{C}^M} \diff^{2M} \balpha_I \int_{\mathbb{C}^M} \diff^{2M} \balpha_S \,P(\balpha_I,\balpha_S) \ket{\balpha_I}\bra{\balpha_I}_I\otimes\ket{\balpha_S} \bra{\balpha_S}_S,
\end{align}
where $\balpha_S = \pars{\alpha_S^{(1)}, \ldots, \alpha_S^{(M)}} \in \mathbb{C}^M$ indexes $M$-mode coherent states $\kets{\balpha_S}$ of $S$, $\balpha_I = \pars{\alpha_I^{(1)}, \ldots, \alpha_I^{(M)}}\in \mathbb{C}^M$ indexes $M$-mode coherent states $\kets{\balpha_I}$ of $I$, and $P\pars{\balpha_I,\balpha_S} \geqslant 0$ is a probability distribution. An average signal energy constraint of  $\cl{N}_S$ implies that $P\pars{\balpha_I,\balpha_S}$ should satisfy
\begin{align} \label{CEC}
\int_{\mathbb{C}^M} \diff^{2M} \balpha_I \int_{\mathbb{C}^M} \diff^{2M} \balpha_S\, P(\balpha_I,\balpha_S) \pars{\sum_{m=0}^M \abs{\alpha_S^{(m)}}^2}  = \cl{N}_S.
\end{align}
The convexity of the QFI \cite{Fuj01}, its invariance under adjoining an idler system in an $\eta$-independent state, and the linearity of the  QFI (\ref{CSQFI}) in the energy then imply that  the QFI of any classical probe Eq.~\eqref{classical probe} obeying the constraint Eq.~\eqref{CEC}  satisfies
\begin{align} \label{QFIcl}
\cl{K}_\eta^{\tsf{cl}} \leqslant \frac{\cl{N}_S}{\eta (2N_B +1)},
\end{align}
which is Eq.~(20) of the main text.

\bibliography{QIarxivbibv2}
\end{document}